\documentclass[journal]{IEEEtran}
\usepackage{mathbbol}
\usepackage{graphicx}
\usepackage{cite}
\usepackage[tight,footnotesize]{subfigure}
\usepackage{url} % \url{my_url_here}
\usepackage[colorlinks = true,linkcolor = blue,urlcolor  = blue,citecolor = blue,anchorcolor = blue]{hyperref}
\let\labelindent\relax
\newcounter{mycount}
\newcommand{\usemycount}{%
	\stepcounter{mycount}%
	\themycount}
	
\begin{document} 

\title{Gamorithm}

\author{Moshe~Sipper and Jason~H.~Moore% <-this % stops a space
\thanks{IEEE Transactions on Games, DOI (identifier) 10.1109/TG.2018.2867743}
\thanks{M. Sipper is with the Institute for Biomedical Informatics (IBI), Perelman School of Medicine,
University of Pennsylvania, Philadelphia, PA 19104 and the Department of Computer Science, Ben-Gurion University, Beer-Sheva 8410501, Israel.}
\thanks{J. H. Moore is with the Institute for Biomedical Informatics (IBI), Perelman School of Medicine,
University of Pennsylvania, Philadelphia, PA 19104.}
\thanks{Corresponding author: M. Sipper (\url{www.moshesipper.com/contact.html}).}
\thanks{Manuscript received \today; revised \today.}}

%\markboth{IEEE Transactions on Games}%
\markboth{Gamorithm, Sipper \& Moore}%
{}

\maketitle

\begin{abstract}
Examining games from a fresh perspective we present the idea of game-inspired and game-based algorithms, dubbed \textit{gamorithms}.
\end{abstract}


\begin{quote}
Le v\'eritable voyage de d\'ecouverte ne consiste pas \`a chercher de nouveaux paysages, mais \`a avoir de nouveaux yeux.

\hfill ---Marcel Proust\footnote{The real voyage of discovery consists not in seeking new landscapes, but in having new eyes.}
\end{quote}

%\bstctlcite{IEEEexample:BSTcontrol} % don't use "----" when same author as before

Applaud.
Beat.
Bet.
Challenge. 
Cheat.
Coach.
Compete.
Defend.
Design.
Draw.
End.
Enjoy. 
Entertain. 
Exhaust.
Fear.
Fight.
Fix. 
Gamble.
Guess.
Hack.
Interact. 
Invent.
Jeopardize.
Kick.
Kill.
Like.
Lose.
Love.
Maneuver.
Manipulate.
Motivate.
Navigate.
Observe. 
Optimize.
Outplay.
Participate. 
Plan.
Play. 
Program.
Qualify.
Quit.
Race.
Risk.
Search.
Solve.
Threaten.
Tie.
Try.
Unravel.
Vie.
Watch. 
Win.
Xpeke.\footnote{Defined by Urban Dictionary as the \textit{action of solo-taking a ``naked Nexus'' in League of Legends game}.}
Yield.
Zoom.

This assortment of seemingly random actions can all be associated with \textit{games}, one of the most ubiquitous of human endeavors. ``Attested as early as 2600 BC, games are a universal part of human experience and present in all cultures.'' \cite{wiki:Game}

Games have been a subject of intense research for decades, both in academia and in industry. 
The field of artificial and computational intelligence (AI/CI) in
games alone admits (at least) 10 broad areas
\cite{lucas2012artificial,yannakakis2015panorama}:
1) nonplayer character (NPC) behavior learning;
2) search and planning;
3) player modeling;
4) games as AI benchmarks;
5) procedural content generation;
6) computational narrative;
7) believable agents;
8) AI-assisted game design;
9) general game AI;
10) AI in commercial games.

And beyond AI/CI there are many other areas of game research: game theory (which models conflict and cooperation between intelligent, rational decision-makers\cite{von1945theory}), social and psychological analysis, historical investigations,\footnote{Of which the recent work by \cite{browne2017}, proposing the study of ancient games as a new frontier for game AI research, is particularly illuminating.} and so forth.

\vspace{5mm}\noindent
``An algorithm is an abstract recipe, prescribing a process that might be carried out by a human, by a computer, or by other means.'' \cite{harel2004algorithmics} Replacing ``algorithm'' with ``game'' in this latter definition underscores the similarity of the two concepts.  
Perchance we might view the area of games under a different light? Specifically, might games not offer us a vast reservoir of potential algorithmic ideas, or \textit{gamorithms}?

The wide diversity of game characteristics offers in turn a wide scope for inspiring novel algorithms. Games can be: 1-, 2-, or $n$-player; discrete or continuous; deterministic or stochastic; played by individuals or by teams; defined tersely (e.g., tic-tac-toe) or wordily (e.g., the game of football, with its 93-page rulebook \cite{nfl-rulebook}). 

``The game is afoot,'' quipped Sherlock Holmes\footnote{And the Earl of Northumberland in Shakespeare's \textit{Henry IV} before him.} who stated even more famously, ``Elementary''---the latter qualifier of which captures the simple essence of our proposed idea:
\begin{quote}
    \textit{To solve a computational problem find or design anew a game-based algorithm.}
\end{quote}

Problem? Gamorithm. 

\vspace{5mm}\noindent
In his seminal paper, ``Computing machinery and intelligence'', Turing \cite{turing1950} devoted a section to a ``Critique of the New Problem'', writing, ``As well as asking, `What is the answer to this new form of the question', one may ask, `Is this new question a worthy one to investigate?''' 
Following this illustrious example we, too, wish to address potential critiques, 
some due to sagacious comments made by the reviewers of the first draft.

(\usemycount) 
\textit{Isn't this simply Serious games?} 
Serious games is an area dealing with games that do not have entertainment as their primary purpose \cite{michael2006,djaouti2011classifying}. These games appear in diverse areas such as healthcare, defense, education, and more. To mention but two examples, Foldit is an online puzzle game about protein folding \cite{khatib2011} and the Google Image Labeler is an image-labeling game \cite{wiki:Google-image}. Serious games is a broad field whereas with gamorithms we wish to focus on algorithmic problem solving.

(\usemycount) 
\textit{What about gamification?} The application of game-design elements and game principles in non-game contexts---\textit{gamification}---is another very broad area \cite{deterding2011game}. This field is probably even broader in scope than serious games, including such disparate cases as Google Local Guides' scoring points and climbing through levels as they upload reviews and photos to Google Maps, and gaming elements in fitness apps (e.g., getting points and badges for various activities). Again, in contrast, gamorithms are far more focused in scope.

(\usemycount) 
\textit{Are gamorithms simulation games?} In a \textit{simulation game}, be it video (e.g., a flight simulator) or real-life (e.g., roleplay), the object is to simulate real-world activities, not solve problems (although, as part of the simulation, problems might be made available for the solving).

(\usemycount) 
\textit{``Why not take an NP-complete problem and transform it into a puzzle?''} This question was posed by \cite{kendall2008} in a survey of NP-complete puzzles.
This is an astute step in the gamorithmic direction, though our net is cast wider, with our interest going beyond NP-complete problems and beyond puzzles. 

(\usemycount) 
\textit{Is there any relation to models in game theory?}
A gamorithm is not a \textit{model} as in game theory, e.g., the iterated prisoner's dilemma that models cooperation between completely ``rational'' individuals \cite{axelrod1984}.

(\usemycount) 
\textit{Alice and Bob come to mind.}
A gamorithm is not a form of ``Alice and Bob'' scenario, which is used in fields such as physics and cryptography for convenience and to aid comprehension \cite{rivest1978method}. 

(\usemycount) 
\textit{The examples of gamorithms provided below are not convincing}. 
This paper resulted from a prolonged period of brainstorming on our part, and we fully acknowledge that we have not yet brought ironclad answers, nor is that our intention. Rather, our aim herein is to raise questions and point to a wealth of possibilities. To put it metaphorically, we perceive this paper to have cracked open a new egg, and appeal to the effervescent games research community to join us in the making of a tasty omelette. While one might pick at this example or that we hope that the totality of them all will serve to stir enough interest and thus pass the baton of brainstorming, as it were.

(\usemycount) 
\textit{This is good-old fashioned algorithmics---what is gained by framing problems as games?} 
Not for naught we began this paper with Proust's adage regarding new eyes. Given that the idea expounded herein is embedded well within the fabric of the field of algorithmics, one can indeed choose to negate its novelty. We feel (and of course this is quite open to debate) that viewing algorithms through gamified glasses offers a beneficent new perspective.

New algorithmic vistas open up when one views a phenomenon, a mathematical theory, a scientific field, or, for that matter, any human or natural endeavor, in a novel way. To mention but a few well-worn examples: considering evolution by natural selection through algorithmic spectacles led to evolutionary algorithms; asking whether ``wet'' neural networks in the brain might be an effective source of inspiration for in silico computing brought forth artificial neural networks; envisioning the direct use of quantum-mechanical phenomena---such as superposition and entanglement---to perform operations on data, gave birth to quantum computing;
questioning the binary nature of Boolean logic resulted in fuzzy logic \cite{sipper02mn}.

(\usemycount) 
\textit{So casting a problem's solution as a gamorithm will help me in some way??} 
Yes, we believe it will, because this casting might provide a possible algorithmic solution (or a path leading to one). The fun factor of games may well motivate research into problems of interest. Moreover, a gamorithmic approach affords us the opportunity to bring the massive amount of research into game-playing and game-solving algorithms to bear. 

Superb algorithms for playing many types of games are now available: board games, card games, dice games, role-playing games, strategy games, video games, first person shooter games, mathematical games, and many more.
A connection forged between a problem and a game might just form a useful bridge. This is somewhat similar to the concept of reduction in computational complexity theory, wherein one transforms one problem into another,  e.g., graph coloring can be reduced to SAT (the satisfiability problem) \cite{garey1979}. Can the recently introduced superb machine Go player \cite{silver2017} serve another purpose and solve a computational problem of interest?
And at a completely different end of the game spectrum, can a massively multiplayer online game like ``World of Warcraft'' solve a computational problem?

\vspace{5mm}\noindent
Having averred what a gamorithm is not and addressed several critiques, 
and bearing in mind that our interest lies in \textit{computing}, \textit{approximating}, and \textit{solving} problems, 
we now provide nine gamorithmic examples by way of proof-of-concept. Note that the problems addressed need not in any way be games or game-related.%\\ 

\setcounter{mycount}{0}

(\usemycount) 
\textsc{Problem:} 
Generate two \textit{random paths} through a map (or graph) with a single cross point. 
% \\
 
\begin{figure*}
\centering

\subfigure[Hex.]{\label{fig-hex}\includegraphics[height=0.20\textheight]{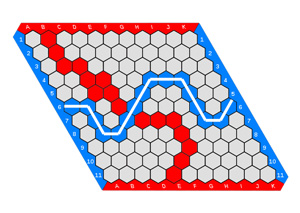}} 
\hfill
\subfigure[Ver Teqs.]{\label{fig-vertex}\includegraphics[height=0.20\textheight]{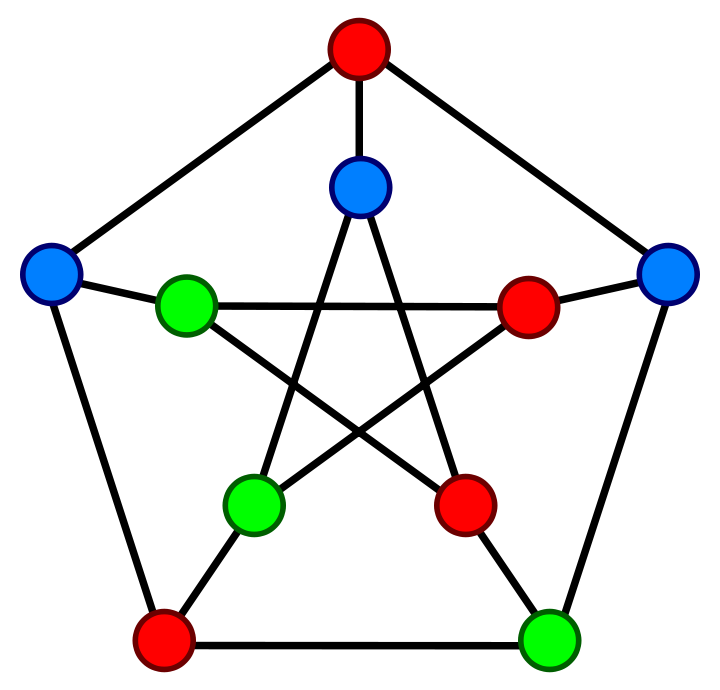}}
\hfill
\subfigure[Sudoku (black---given values, red---solution).]{\label{fig-sudoku}\includegraphics[height=0.20\textheight]{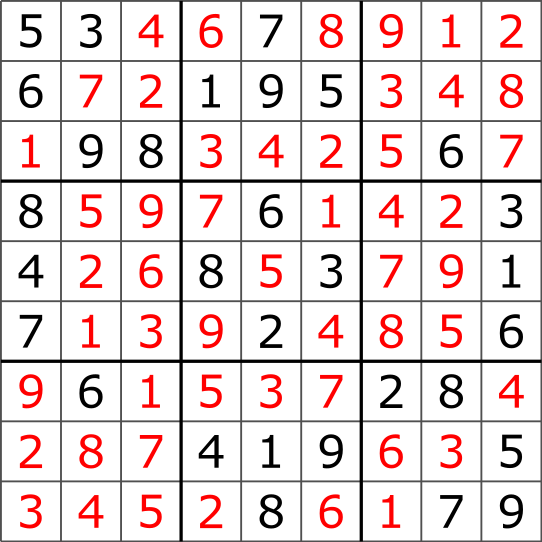}} 

\subfigure[Tetris.]{\label{fig-tetris}\includegraphics[height=0.20\textheight]{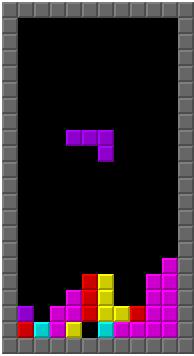}}
\hfill
\subfigure[Rush Hour.]{\label{fig-rushhour}\includegraphics[height=0.20\textheight]{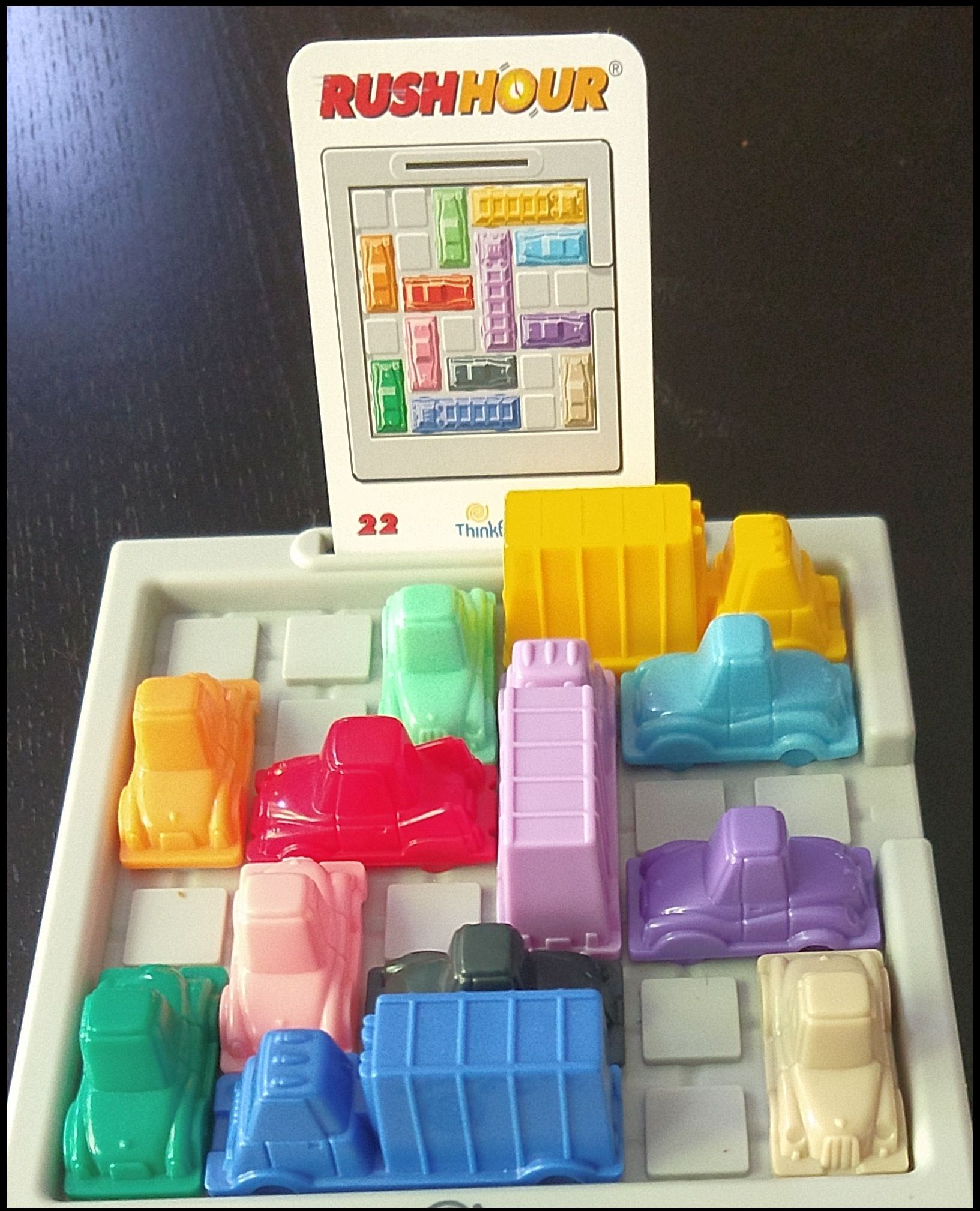}}
\hfill
\subfigure[Tennis]{\label{fig-tennis}\includegraphics[height=0.20\textheight]{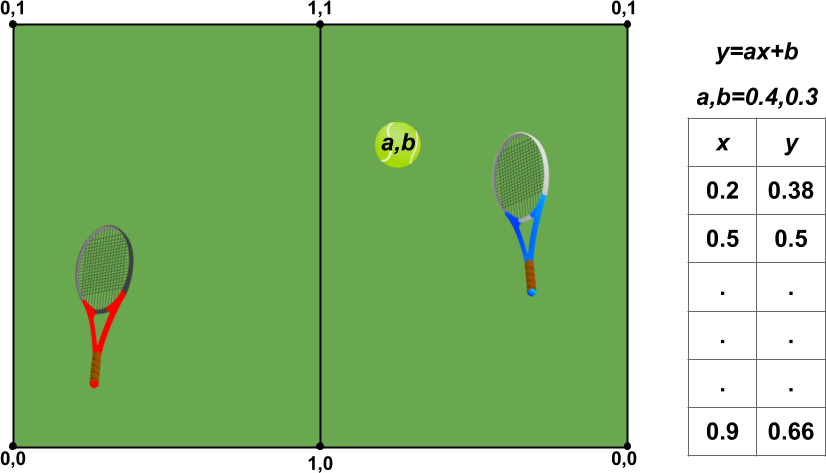}} 

\subfigure[Jenga.]{\label{fig-jenga}\includegraphics[height=0.20\textheight]{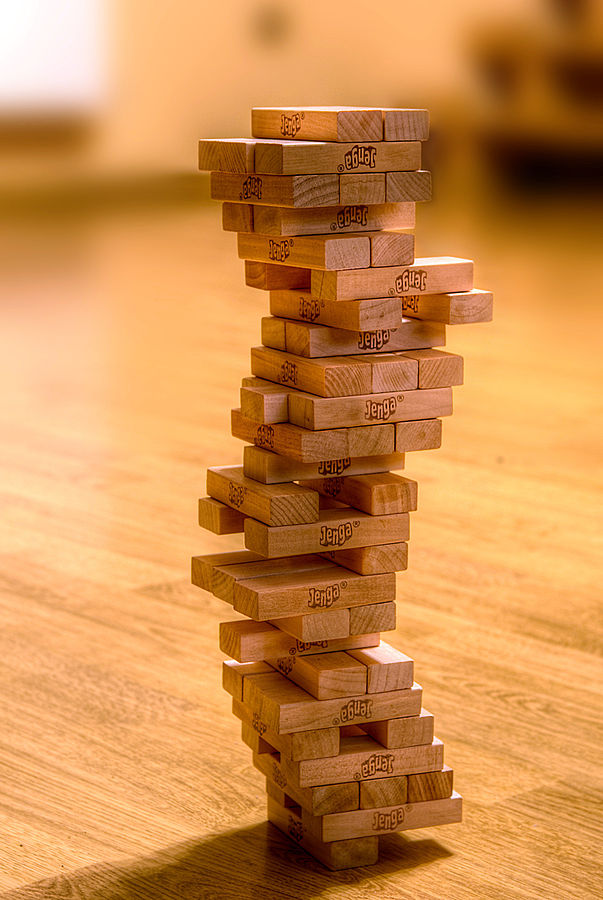}}
\hfill
\subfigure[Monopoly.]{\label{fig-monopoly}\includegraphics[height=0.20\textheight]{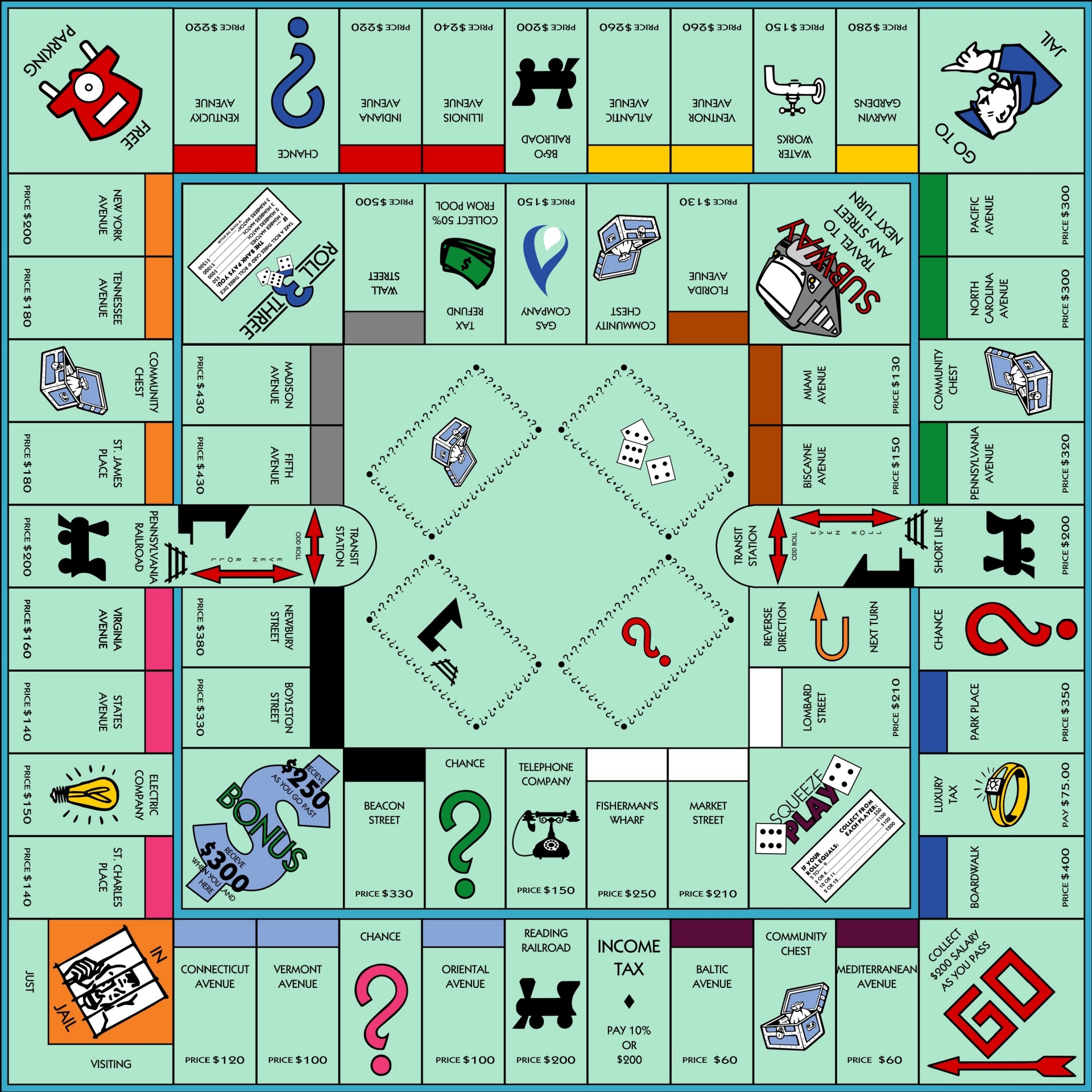}}
\hfill
\subfigure[Ackley's function as a Minecraft world?]{\label{fig-ackley}\includegraphics[height=0.20\textheight]{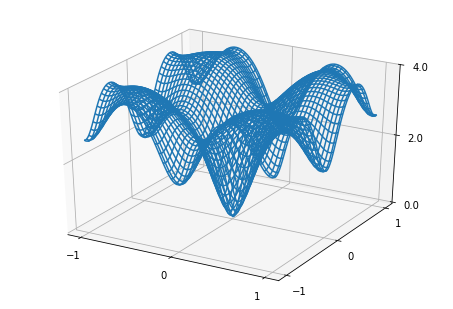}} 

\caption{Gamorithms.\newline
Image sources:
(a) \url{https://en.wikipedia.org/wiki/File:Hex-board-11x11-(2).jpg},  
(b) \url{https://commons.wikimedia.org/wiki/File:Petersen_graph_3-coloring.svg},
(c) \url{https://commons.wikimedia.org/wiki/File:Sudoku_Puzzle_by_L2G-20050714_solution_standardized_layout.svg},
(d) \url{https://commons.wikimedia.org/wiki/File:Tetris.jpg},
(e) photo by authors,
(f) artwork by authors,
(g) \url{https://commons.wikimedia.org/wiki/File:Jenga_distorted.jpg},
(h) \url{http://www.publicdomainpictures.net/view-image.php?image=203219&picture=monopoly-game-board}
(i) artwork by authors.
}

\label{fig-games}
\end{figure*}

\textsc{Gamorithm:} 
\textit{Hex}, a 2-player, strategy board game, where players alternate placing pieces on unoccupied spaces of a board, attempting to link their opposite sides in an unbroken chain (Figure~\ref{fig-hex}). Graph problems in general may be well suited to connection games \cite{wiki:Connection-game}, a category of which Hex is a prominent member. 

%$\quad$
(\usemycount) 
\textsc{Problem:} 
\textit{Graph coloring} is an assignment of labels (``colors'') to elements of a graph subject to certain constraints. For example, in the vertex-coloring problem each vertex must be assigned a color such that no two adjacent vertices (i.e., with a common edge) share the same color. This problem is usually NP-complete, although some special cases are polynomial-time \cite{malaguti2010survey}.
% \\

\textsc{Gamorithm:} 
In the 1950s Claude Shannon invented an abstract strategy game for two players, known as the \textit{Shannon switching game}. The game is played on a finite graph between two alternating players, \textit{Cut}  and \textit{Short}, the former deleting a non-colored edge in her turn, the latter coloring any edge still left in his turn. There are two special nodes, $A$ and $B$, where Cut wins if she turns the graph into one where $A$ and $B$ are no longer connected, and Short wins if he manages to create a colored path from $A$ to $B$.
An explicit solution was found in 1964 \cite{lehman1964}.

We might invent new games on graphs, such as a ``relative'' of the switching game, dubbed \textit{Ver Teqs}, wherein two players alternately place colored pieces on the graph, respecting the no-adjacent-same-color rule (Figure~\ref{fig-vertex}). A player who finds herself in a position where she must break the rule---loses. If all vertices are colored legally, a tie is reached---as well as a solution to our problem. Note that even a game that does not reach a tie may still provide an acceptable, approximate solution.

%$\quad$
(\usemycount)
\textsc{Problem:} 
\textit{Imputation of missing data} in a table of data values \cite{gelman_hill_2006}, given various constraints on placement within rows, columns, and specific regions.
% \\

\textsc{Gamorithm:} 
Forms of \textit{Latin square} \cite{wiki:Latin-square} games come to mind, one prime example being \textit{Sudoku} (Figure~\ref{fig-sudoku}).

%$\quad$
(\usemycount)
\textsc{Problem:}
\textit{Packing problems} are a class of optimization problems that involve attempting to pack objects together into bins or containers \cite{dyckhoff1990}. 
Usually, the goal is either  to pack a single container as compactly as possible or pack all objects using as few containers as possible.
In dynamic problems, objects arrive over time and repacking may or may not be allowed \cite{coffman1983}. 
% A well-known example is 
% In the \textit{bin-packing problem}, objects of different volumes must be packed into a finite number of bins or containers, each of a given volume, in a way that minimizes the number of bins used. 
% \\

\textsc{Gamorithm:}
We propsose \textit{Tetris}-like games (Figure~\ref{fig-tetris}) as gamorithms for solving dynamic packing problems with no repacking \cite{dyckhoff1990,berndt2014,gupta2017}.

%$\quad$
(\usemycount)
\textsc{Problem:} 
With the age of ubiquitous autonomous vehicles well-nigh upon us, one might imagine giant parking lots (perhaps at city edges), where self-driving cars and trucks plunk down at night. This may well engender interesting problems of \textit{compact packing} and \textit{complex routing}. For example, what if a need for a specific car arises, which must thereupon make its way through the mass of parked vehicles all the way to the exit? 
% \\

\textsc{Gamorithm:}
\textit{Rush Hour} is a board game that asks precisely that question, with computational intelligence solutions to boot 
\cite{hauptman2009gp,Sipper2011Win} (Figure~\ref{fig-rushhour}).

%$\quad$
(\usemycount) 
\textsc{Problem:}
\textit{Polynomial regression}, a common problem in which the relationship between the independent variable $x$ and the dependent variable $y$ is modelled as an $n$th degree polynomial in $x$.
% \\

\textsc{Gamorithm:} 
A form of \textit{tennis} match can be adapted to serve as a gamorithm for this problem. 
Consider the example in Figure~\ref{fig-tennis}, where we are given a table of independent and dependent variables, drawn from the polynomial $y=ax+b$, with $a=0.4$ and $b=0.3$. The goal is to find $a$ and $b$. 

We conduct a tennis match in the search space of
$a,b\in[0,1]$, where the ball represents a pair of $\{a,b\}$ values. Each of the two player's sides of the court is a plane representing the search space. 
The quality of a shot is calculated when
the ball lands in a player's court, by means of a specified cost function (mean absolute error, root mean squared error, etc'). 
The mechanics of the game are handled by a  
tennis controller (simulator), whose dynamics can be as simple or as complex as we wish. At the simple end an
elementary formula might be used to calculate a player's response strike in terms of, say, speed and angle, using the shot's quality alone; at the complex end one might implement full-blown physics, with sophisticated player strategies that use more information and memory to calculate a strike. 

The tennis gamorithm can be generalized in any number of ways, e.g., by adding court dimensions (thus increasing the polynomial degree), by adding players, and by adding nets (i.e., the court is not divided into two halves but into $n$ partitions, whereupon new playing rules need to be defined).
Interestingly, we are not concerned herein with which player wins the game but rather with having both players cooperate through competition to solve a problem. Essentially, the best shot in the game (i.e., lowest cost-function value) is our trophy. 

%$\quad$
(\usemycount)
\textsc{Problem:}
Given an undirected graph $G(V, E)$, a \textit{global min-cut} is a partition of $V$ into two subsets $(A, B)$ such
that the number of edges between $A$ and $B$ is minimized \cite{karger93}. 
% \\

\textsc{Gamorithm:}
In \textit{Jenga}, players take turns removing one block at a time from a tower constructed of 54 blocks. Each block removed is then placed on top of the tower, creating a progressively taller and unstable structure
(Figure~\ref{fig-jenga}).
The game ends when the tower falls, or if any piece falls from the tower other than the piece being removed to move to the top. The winner is the last person to successfully remove and place a block.

Jenga-like games may well fit the gamorithmic bill where min-cut problems are concerned, with the objective being to partition graphs---or, more generally, multi-piece objects---with a minimal number of operations, cuts, or moves. (There are various other kinds of cut problems, e.g., in minimum $k$-cut one seeks a set of edges whose removal would partition the graph into $k$ connected components.)

%$\quad$
(\usemycount)
\textsc{Problem:}
\textit{Facility location problems}, studied in operations research and computational geometry,  are concerned with the optimal placement of facilities to minimize transportation costs while considering factors like avoiding placing hazardous materials near housing, and the facilities of competitors \cite{guha1999}. 

\textsc{Gamorithm:}
\textit{Monopoly} is a popular board game where players
move at random (based on a dice roll), developing and selling properties on a game board (Figure~\ref{fig-monopoly}). 
The game shares some basic commonalities with facility location and might be adapted to form a gamorithm, by customizing the rules and board such that players compete to attain a tenable solution to the original problem. 
This design need not be as daunting a task as one might think. As is often the case with computational problems, we can begin with a simple scenario (e.g., a small number of facilities and basic rules) and gradually work our way up to a more complex game.

%$\quad$
(\usemycount) 
\textsc{Problem:} 
More of a meta-problem, we propose the use of \textit{virtual-world} games (e.g., video games) as problem solvers.

\textsc{Gamorithm:}
With virtual worlds one usually relies on extensive 
algorithmic creation of game content to build and populate the scene, so-called \textit{procedural content generation} (PCG) \cite{shaker2016}. What if, instead, the content represented the search space of a problem of interest, e.g., a complex, real-valued optimization problem \cite{liang2014} (think of playing Minecraft in the world of Figure~\ref{fig-ackley})? Might  
there be a beneficial commonality between making one's way (intelligently) through a virtual world and searching (intelligently) through a search space?

%$\quad$

\vspace{5mm}\noindent
Having quoted Alan Turing earlier, it is perhaps fitting to end with Turing-Award winner Judea Pearl, who recently wrote: ``One final comment about these `games' ... they are quite obviously not games but
serious business. I have referred to them as games because the joy of being able to solve them swiftly and meaningfully is akin to the pleasure a child feels on figuring out that he can crack puzzles that stumped him before. Few moments in a scientific career are as satisfying as taking a problem that has puzzled and confused generations of predecessors and reducing it to a straightforward game or algorithm.'' \cite{pearl2018} 

We believe that virtually any game has the potential of leading to a gamorithm that will solve some problem or other.  The list of games available for examination is quite large, so much so in fact that Wikipedia's game-list entry is actually a list \textit{of lists} \cite{wiki:Lists-of-games}.
And, fulsome as it is, this list contains but extant games, leaving an infinitude of new games---and gamorithms---yet to be imagined.

\vspace{15pt}
\begin{quote}
``Prepare for unforeseen consequences.''

\hfill ---\textit{Half-Life 2: Episode Two} (video game)
\end{quote}
\vspace{5pt}

\section*{Acknowledgment}
We are grateful to the anonymous reviewers for their helpful comments.

This work was supported by National Institutes of Health (USA) grants LM010098 and AI116794.

\clearpage

\end{document}